# Adobe AIR, Bringing Rich Internet Applications to the Desktop

**Valentin Vieriu, Cătălin Țuican**
**"Tibiscus" University of Timişoara, România**

ABSTRACT. Rich Internet Applications are the new trend in software development today. Adobe AIR offers the possibility to create cross-platform desktop applications using popular Web technologies like HTML, JavaScript, Flash and Flex.
This article is focused on presenting the advantages that this new environment has to offer for the web development community and how quickly you can develop a desktop application using Adobe AIR.

## 1. Defining a rich Internet application

A rich internet application (RIA) combines the best features of a desktop application with broad reach, low cost of deployment of web application.

In the early days of software development, big and complex applications were developed on powerful servers and were accessed by the user using a terminal connection.

When the client computer started to become smarter, a new type of programming paradigm appeared. The programs running on the client machine offered a more appealing interface, and advanced functionalities. Also this software could take advantage of the continue increasing power of the desktop computers and not base its strategy only on the server processing power. This approach also brought some major disadvantages, one of them being the heavy maintenance process. Every time a change in the application occurred, every single client machine had to be updated. This soon became a major drawback.

367

When the web applications appeared, we saw a step backward in functionality because most of the processing was done on the server, and the client machine running a web browser became a dumb terminal.

The new rich internet applications have all the benefits of the client/server type of development but none of the maintenance problems because they are delivered using a proprietary web browser plug-in or independently via virtual machines.

RIA's are enhanced Web applications that have additional functionality on the client, which makes them more responsive to the user than standard HTML pages. RIA technologies include AJAX, which is a combination of JavaScript, dynamic HTML and an asynchronous server request interface; Flash, a widely available plug-in technology from Adobe often used by designers; Flex, a variation on Flash more suited to programmers; Silverlight, a relatively new plug-in technology from Microsoft that includes a subset of the .Net Framework; and Curl, an object-oriented language with embedded HTML markup.

RIA's bring a lot Expressiveness to the user. His experience with an online application does not need to be dull. He wants to see smooth menu animations, drop shadows, vector graphics, and maps manipulation. Rich media content like audio and video and real time data pooling and messaging are common now inside RIA's.

In this article we will talk about how Adobe succeeded to bring rich Internet applications to the Desktop and also how easy is for a web developer to transform a web application to a desktop one.

## 2. What is Adobe AIR

Adobe Integrated Runtime (AIR) is a cross-platform runtime environment for building rich Internet applications using Adobe Flash, Adobe Flex, HTML, or Ajax, that can be deployed as a desktop application.

AIR, the Adobe integrated runtime from Adobe systems, has been in developments from the last couple of years. Developers are now able to build cross platforms applications, leveraging their existing skills in HTML, Flash and Flex and deploying them on almost all operating systems.

Using AIR developers can create applications that combine benefits of web application like: network and user connectivity, rich media content, easy development and broad reach with the strengths of the desktop applications like : interaction with other applications, local resource access, offline access to information, rich interactive experience.

Adobe AIR it's not another programming language, it's just a runtime that as it allows existing Flash, Actionscript or HTML and JavaScript code to be used to construct a more traditional desktop-like program. For Flex developers anything you can do in Flex, you can do in an AIR application built in Flex and the same thing is for Flash or HTML.

Adobe AIR is coming from the Web to the desktop and is targeted at web developers. Its primary use case is to enable web applications and RIA's to be deployed to the desktop.

The runtime itself is installed as a software on a number of different operating systems and it supports the installation and running of desktop applications that are built in the Flex, Flash and HTML architectures. There's a whole set of development tools that are provided by Adobe for building these applications. Some of these development tools are free and some of them are incorporated into existing commercial tools such as Flex Builder, Flash and Dreamweaver.

The Adobe AIR runtime may be a relatively new platform, but it actually embeds three highly mature and stable cross-platform technologies to power AIR applications. These are the following:

**WebKit**: Used for rendering HTML content inside an AIR app. WebKit is an open source, cross-platform browser and is the base layer for Apple's Safari browser. WebKit is known for its strong support of W3C standards, such as HTML,XHTML, Document Object Model (DOM), Cascading Style Sheets (CSS), and ECMAScript. However, it also provides support for enhanced functionality (enabling the creation of rounded corners using CSS). Because developing for AIR means developing for WebKit, you're free to take advantage of these nonstandard extensions and not worry about browser compatibility.

**Adobe Flash Player**: Used for playing Flash media (SWF files). Flash Player is a cross-platform virtual machine used to run media created in the Adobe Flash authoring environment and full SWF-based applications created using Adobe Flex.

**SQLite**: A database engine for enabling local database access. It's an extremely lightweight, open source, cross-platform SQL database engine that is embedded in many desktop and mobile products. In contrast to most SQL databases, it doesn't require a separate server process, and it uses a standard file to store an entire database.

AIR applications can be installed and run on a number of different operating systems. These include Windows, all editions, and Macintosh, on the latest versions of the Mac OSX operating system. After a few months from the releasing, Adobe AIR was also able to running on Linux operating systems.

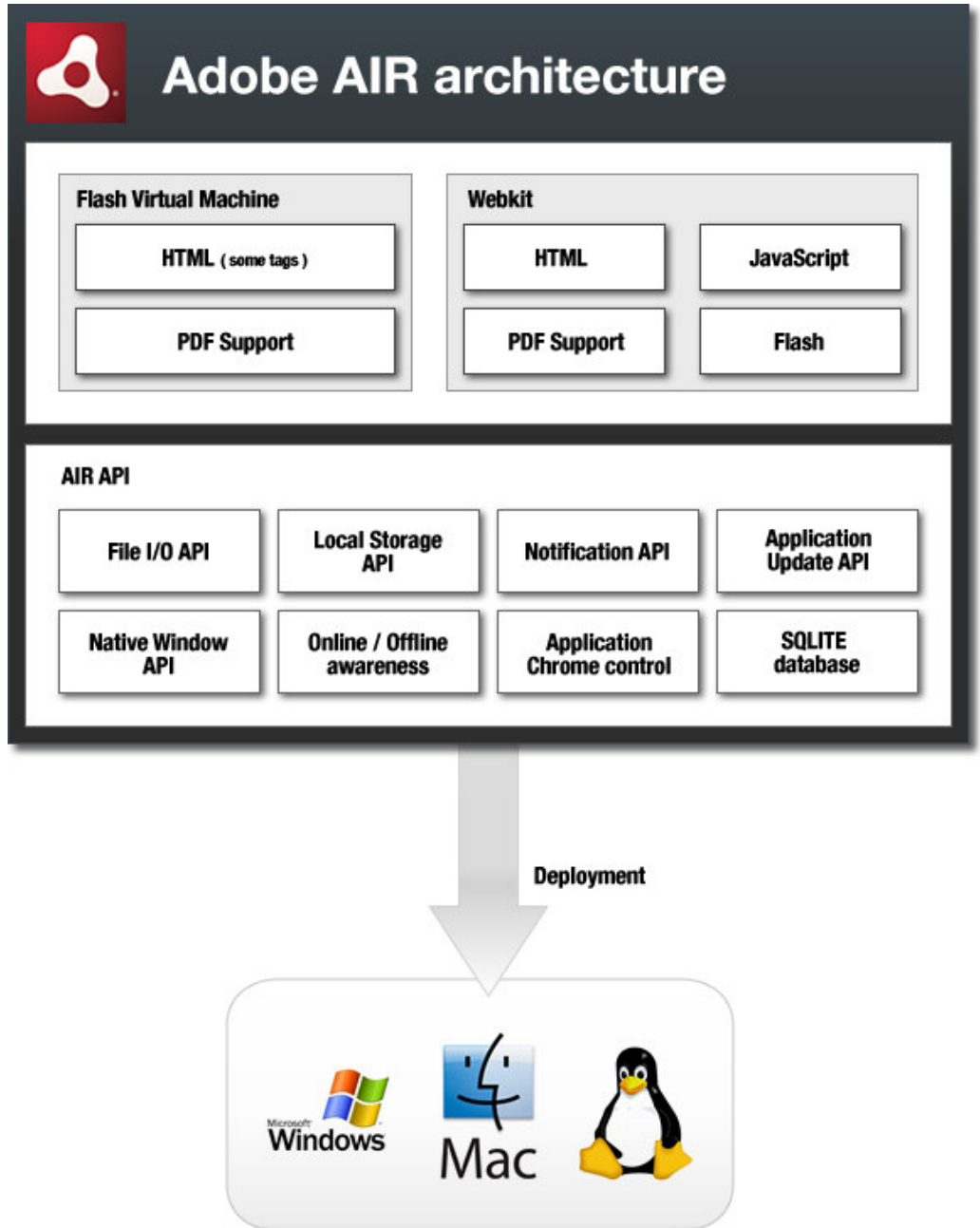

Fig.1 Adobe AIR architecture

### 3. Advantage of the AIR

Looking at the core AIR architecture we see Adobe Flash Player and WebKit, sitting on top and tightly communicating with the AIR API (Application Program Interface). This API offers AIR applications the next advantages:

- the possibility to save and load files on the user computer,
- automatically update the application when the user goes online
- native window support with usual operating system functionalities (maximize, minimize, system tray icon, always on top option)
- advanced chrome control with transparent windows, transparent video
- a notification system that allows communications between air applications running on the same computer
- a local SQL database system which use a standard file to store the information

All of the platforms and languages that can be used in building AIR applications have existed for quite a while and are completely documented and there's a lot of support out there from Adobe. For instance, if you're Flex developer you'll be working with MXML and ActionScript 3. If you're Flash developer, you'll be working with ActionScript. And if you're an HTML developer, you'll be working with HTML, cascading style sheets and JavaScript. The bottom line is that there aren't new languages to learn so much as new techniques and packaging them and leveraging the capabilities of the runtime.

Having Flash Player and WebKit rendering engine integrated inside AIR opens many possibilities for AIR developers. An AIR app can consist of several combinations:

- HTML/ JavaScript only
- HTML and Ajax
- Flash only
- Flex only
- Flash/Flex and Html

Also take into consideration that:

- you can access the Flash Player and ActionScript Library API's from JavaScript
- you can call JavaScript and access and modify HTML DOM from inside Flash/Flex

You can see that Adobe AIR breaks down the traditional walls that have existed in Web development architecture.

371

The users have access to complete Web services, including REST-based services, meaning XML that can be retrieve from a Web server at runtime, SOAP-based Web services that leverage the SOAP protocol and Flash remoting-based services. Adobe integrated runtime have different software development tools that can be select from to build, compile and deploy AIR applications.

## 4. Software Development Tools

There's a command line compiler, for Flex developers, called amxmlc. This is the same to the MXML C command line compiler that's a part of the Flex SDK. With MXML and ActionScript 3 assets, users can build Flash-based content using this command line compiler. There's another command line compiler called acompc, which is the same to the Flex compc command line compiler, and can be used to build components and component libraries.

In the software developer kit (SDK) there's a debug launcher that can be use to debug and test applications, and then the important piece of the SDK is the AIR developer tools or the Adobe developer tools (ADT). ADT can export complete installation files. An AIR installation file it's built using the ADT tool and has the file extension named .air. If we take Flash and HTML formatted content and combines that with something called an application descriptor file, we can build an AIR application and from there can be build the AIR install file. The ADT also create something called a self-signed security certificate. We will get from a main security vendor a security certificate, but if we build an application for internal use, the ADT allows us to build lower level security certificates that don't require a purchase from a security vendor. Integration with Apache ANT package is supported and documented.

A set of commercial development tools is also coming with Adobe AIR.

A new version of Flex Builder has been released, which have integrated support for building and Flex-based AIR applications. From the start, when is created a new Flex project, there are two options: creating a web-based project or a desktop project. And when is selected the desktop project, will be done an AIR application. All of the command line tools that are available in the SDK are also integrated into Flex Builder. And if users have Flex Builder, it makes it a lot easier to build these applications.

For Flash developers there's an upgrade to a newer version that is needed. That's a free upgrade to the existing Flash CS3 package, so its not needed to purchase anything else. And then there's an additional patch that needs to be installed that adds commands to the Flash, to support creating Flash-based AIR applications directly from that environment.

Those are the major development tools from Adobe, the free SDK for building AIR applications from the command line and the commercial development tools, the new Flex Builder 3 and upgrades to Flash. Whether you're a Flex, Flash or HTML developer, Adobe delivers these tools to help you get started building AIR applications.

## 5. Security

There are two aspects of security in AIR:
- installation security
- runtime security

In both cases the security model is determined by the fact that AIR applications are desktop applications, and just like applications such as Notepad or other application that you install locally, an AIR application has the ability to access local resources, such as the local file system, printers etc.

Web applications, built in HTML, Flash or Flex, have to run within the sandbox. This means that as content is downloaded from the internet, that content, can't access local resources (content like Flex applications, Flash-based content or scripting routines)

The problem with the Web is that users freely download content from the web and just blindly trust it and the world of the web browser has been carefully designed to make sure that content coming from the Internet doesn't have the ability to disrupt or damage content on the local system. When a user installs an application as a desktop application, they grant much broader access to their system. And so there are two aspects to installation of these applications that have to be handled. During the installation process the user has to be given adequate information that allows them to make an informed decision as to whether to grant this local access. At run time, the application has to make informed decisions as to which of its components will be allowed to access content on the desktop and which won't. In order to use this access, AIR applications have the ability to read and write to the local disk simple text files and complex data. This is implemented through a set of code libraries that can be use in access in AIR applications. There's a set of ActionScript libraries that are use from both Flash and Flex and a set of job descript base libraries to use in HTML

373

applications. These libraries can read and write text files to the local disk and then in addition there's data access.

Security in AIR installations is managed through signing. When is created an AIR application, this application must provide something called a digital certificate. Digital certificates verify the identity. For creating digital certificates are two ways. All of the development tools, Flex Builder and Flash, once they've been upgraded to incorporate the AIR development tools, allow us to create our own self-signed certificates. These certificates are created without any cost and they're very easy to create, but they're only designed to be used for internal or for low security environments. This is because when is installed an AIR application that's been signed with the self-signed certificate the user sees that publisher is unverified.

At runtime, security is implemented again around the security sandbox. The application sandbox makes a distinction between resources that are on the internet and resources on the client system. Any content that the user has explicitly installed has both complete access to the client system and internet-based services.

## 6. Creating a small AIR application

The easiest way to develop AIR applications is to use an IDE (Integrated Development Environment). Adobe offers here a lot of proprietary software: Adobe Flash, Adobe Flex, Adobe Dreamweaver. We will use Aptana Studio (http://www.aptana.com/air/ ) a free open source cross platform IDE for building Ajax web applications. It includes support for JavaScript, HTML, DOM, and CSS with code-completion, outlining, JavaScript debugging, error and warning notification and integrated documentation. Additional plugins allow Aptana Studio to be extended to support Ruby on Rails, PHP, Python, Perl, Adobe AIR, and Apple iPhone.

After downloading Aptana and following the guide at http://www.aptana.com/air/ we will need to install Adobe AIR runtime (http://www.adobe.com/go/getair). The runtime is necessary on every computer that wants to run AIR applications.

We will work in a project environment, so we need to create a new project. File -> New -> Project and choose Adobe AIR Project.

The next wizard will generate de application.xml file that will tell the AIR compiler important information about your application. This XML can be edited manually later if you wish to change something.

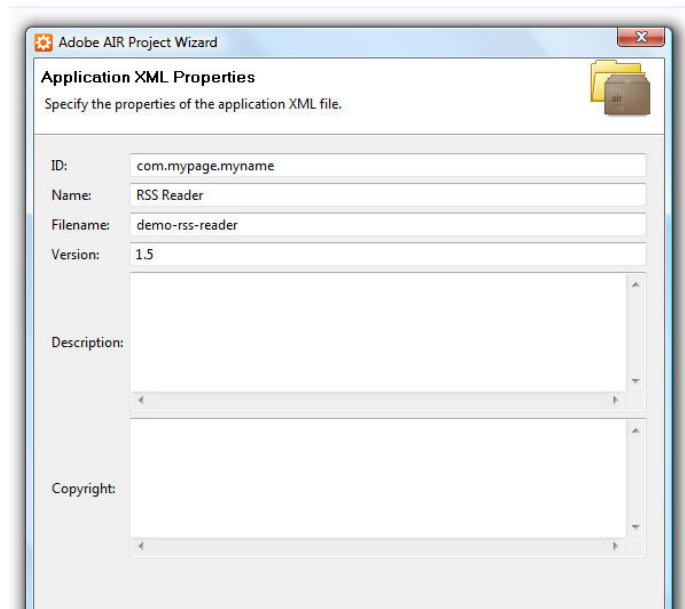

Fig.2 Creating an AIR project inside Aptana

Every field here is self-explanatory. I will mention some of the fields:
Name- it's the name of the project
Filename – it's the filename the application will have after compilation
Version – you can give your file a version to keep track of the changes
ID – this is an unique identifier, usually here you enter the your reverse domain name.

If you have a page *myname.mypage.com*, this ID can be *com.mypage.myname*

In the next screen you can set some restrictions to the size of the application window and also here you can decide if you want to use the system chrome (the classic mode of drawing windows. It's handled by the operating system and will look different depending on your operating system). If you want a custom window, with your own design, you will choose Custom Chrome. (Fig 3)

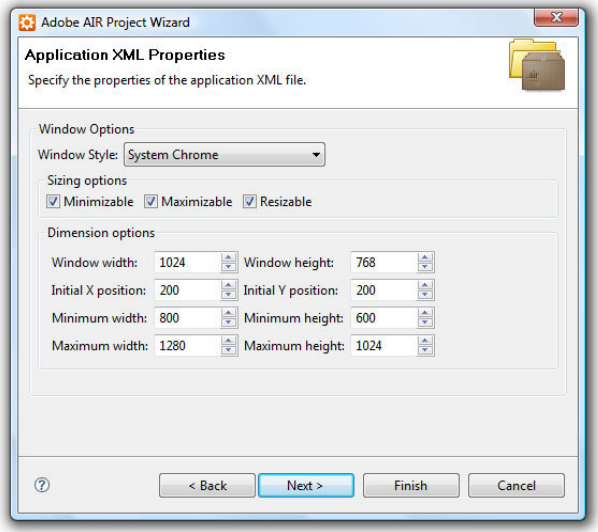

Fig.3 Setting the Chrome inside Aptana

In the last steps, Aptana offers you the possibility to include various JavaScript libraries and also some JavaScript tools for easy debugging. We will use jQuery, a very popular library, so we will check that library so it will be automatically included.

As I said before the role of this wizard is to create an XML file. This is the configuration file that th Adobe AIR SDK takes into consideration when compiling the application. If you need to change something from the previous steps, you can do that by simply editing the application.xml file.

We will create a simple application that will gather a news feed (RSS feed). The URL of the RSS feed it will be extracted from the computer Clipboard.

An RSS feed is actually an XML file, that it's been standardized and now it's being used for content and news sharing. Here are some of the required tags for an RSS feed. We will use those tags later to extract the information that we need.

- **rss**. The global container.
- **channel**. A distributing channel. It has several descriptive tags and holds one or several items.
  **Required tags for the channel**
- **title**. The title of the channel. Should contains the name.
- **link**. URL of the website that provides this channel.
- **description**. Summary of what the provider is.

- one item tag at least, for the content

**Items of the channel**

- **title**. Title of the article.
- **link**. The URL of the page.
- **description**. Summary of the article..

```xml
1  <rss version="2.0">
2  <channel>
3      <title>Xul</title>
4      <link>http://www.xul.fr/</link>
5      <description></description>
6      <item>
7          <title>Xul news</title>
8          <link>http://www.xul.fr/en-xml-rss.html</link>
9          <description>... some text... </description>
10     </item>
11 </channel>
12 </rss>
```

Fig.4 A minimum required RSS feed XML tags

Getting back to our application here is how the main HTML file looks:

```html
<html>
    <head>
        <title>Simple RSS feed reader</title>
        <script type="text/javascript" src="jquery.js"></script>
        <script type="text/javascript" src="app.js"></script>
        <script type="text/javascript" src="lib/air/AIRAliases.js"></script>
    </head>
    <body>
        <h1>Latest News</h1>
        <div id="feedContent">
        </div>
    </body>
</html>
```

Fig. 5

It is a very simple HTML file. We import 3 JavaScript files. One is jQuery, a very popular framework, another is AIRAliases.js, this file will offer us easy access to AIR API's. It is included in the AIR SDK. And our custom app.js file where we will write our application. Here is the first content of that file:

377

```
1  function get_rss_feed(rssfeed){
2      //clear the content in the div for the next feed.
3      $("#feedContent").empty();
4      //use the jQuery get to grab the RSS feed from the internet
5      $.get(rssfeed, function(data){
6          //find each 'item' in the file and parse it
7          $(data).find('item').each(function(){
8              //we create a loop for parsing all the items inside the RSS feed
9              var $item = $(this);
10             // grab the post title
11             var title = $(this).find('title').text();
12             // grab the post's URL
13             var link = $item.find('link').text();
14             // next, the description
15             var description = $item.find('description').text();
16             //don't forget the pubdate
17             var pubDate = $item.find('pubDate').text();
18
19             // 'html' will contain the information that we will insert in the HTML page
20             var html = "<div class=\"entry\"><h2 class=\"postTitle\">" + title + "</\h2>";
21             html += "<em class=\"date\">" + pubDate + "</em>";
22             html += "<p class=\"description\">" + description + "</p>";
23             html += "<a href=\"" + link + "\" target=\"_blank\">Read More >></\a></div>";
24
25             //put that feed content on the screen!
26             $('#feedContent').append($(html));
27         });
28     });
29  };
30  //We use the document ready approach to make sure that the page is fully loaded before
31  //launching the JavaScript code
32  $(document).ready(function(){
33      var clipboardContent = air.Clipboard.generalClipboard.getData(air.ClipboardFormats.TEXT_FORMAT);
34      if (clipboardContent == null) {
35          alert('Please copy a RSS feed link to the clipboard.');
36      }
37      else {
38          get_rss_feed(clipboardContent);
39      }
40  });
```

Fig. 6

We use $(document).ready() to make sure that everything is loaded before we start our JavaScript.

Getting the content from the Clipboard is very simple. Using AIRAliases.js we have access to AIR API's and get that URL from the clipboard. ( see line 33-39 ).

Once we have the URL we can get the RSS feed from online. We use jQuery for parsing the result, which is actually an XML file with some general established tags. We extract the information we need from the XML ( line 7-18 ) and after we format the code ( line 22 24 ) we insert it into the HTML file ( line 26 ).

Before running the application we go to any news provider website ( I choose http://ajaxian.com ) and find the link to the RSS feed, right click and choose copy link (http://feeds2.feedburner.com/ajaxian).

We click run inside Aptana 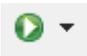 and, because we have the URL of the RSS feed in our clipboard, our AIR application will extract that information and retrieve the content from the web and render it on the screen.

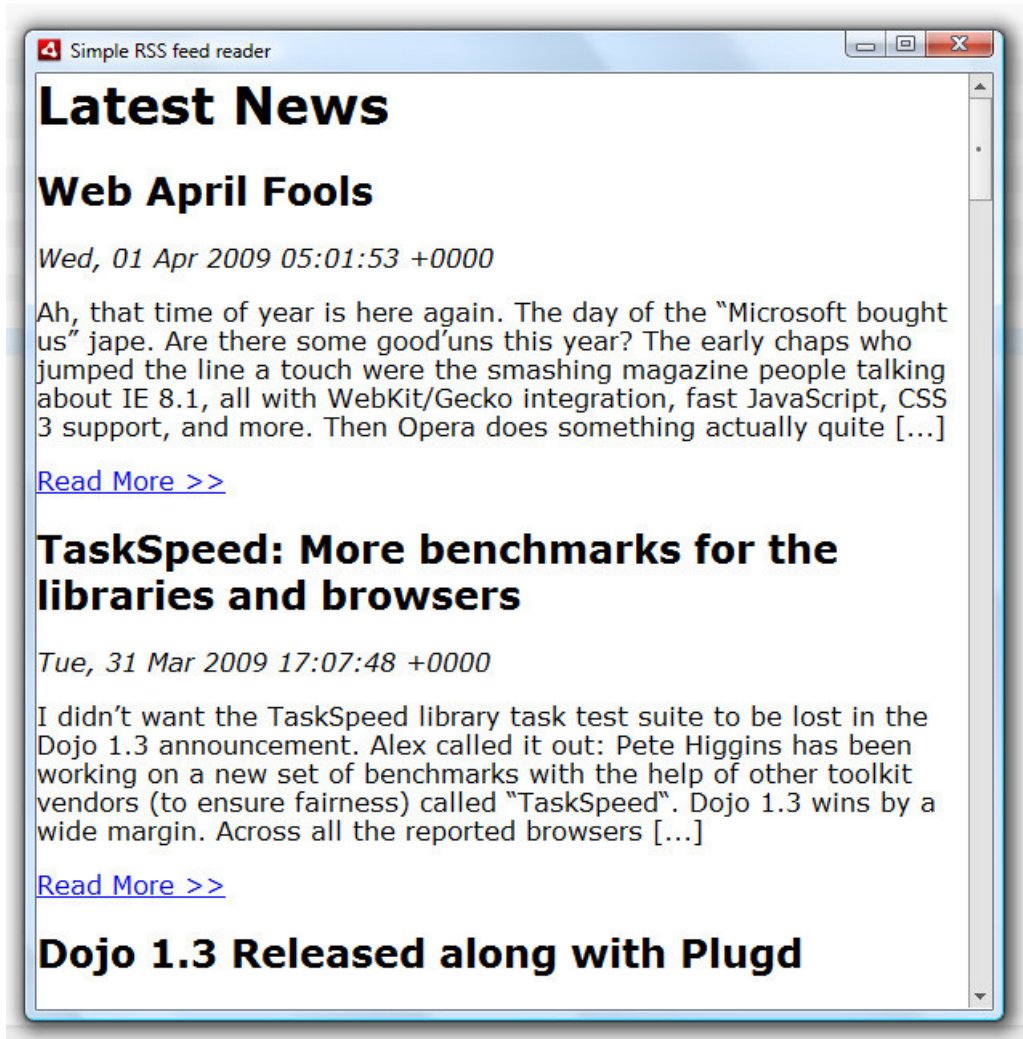

Fig. 7

## Conclusions

Even if this is a simple application we can see that using simple Web Development skills we can develop in a very short amount of time a desktop application that will run on most operating systems today.

This example shows how easy it is to create a desktop application without using proprietary software and expensive licenses.

You can use free and open source development tools like Aptana, simple and easy to learn technologies like HTML, JavaScript and CSS and create cross platform desktop applications.